\documentclass[preprint]{emulateapj}
\usepackage{hyperref}
\usepackage[usenames,dvipsnames]{color}
\hypersetup{colorlinks,citecolor=Blue,linkcolor=Red,urlcolor=Blue}
\usepackage[titletoc]{appendix}
\usepackage{amsmath}

\begin{document} 

\title{Stellar winds as a mechanism to tilt the spin axes of Sun-like stars}

\author{Christopher Spalding$^{1}$} 
\affil{$^1$Department of Astronomy, Yale University, New Haven, CT 06511} 
\begin{abstract}
The rotation axis of the Sun is misaligned from the mean angular momentum plane of the Solar system by about 6 degrees. This obliquity significantly exceeds the $\sim1-2$\,degree distribution of inclinations among the planetary orbits and therefore requires a physical explanation. In concert, Sun-like stars are known to spin down by an order of magnitude throughout their lifetimes. This spin-down is driven by the stellar wind, which carries angular momentum from the star. If the mean angular momentum axis of the stellar wind deviates from that of the stellar spin axis, it will lead to a component of the spin-down torque that acts to tilt the star. Here, we show that Solar-like tilts of 6 degrees naturally arise during the first 10-100\,Myr after planet formation as a result of stellar winds that deviate by about 10 degrees from the star's spin axis. These results apply to the idealized case of a dipole field, mildly inclined to the spin axis. Time-variability in the misalignment between the magnetic and spin poles is modeled as stochastic fluctuations, autocorrelated over timescales comparable to the primordial spin-down time of several million years. In addition to wind direction, time-variability in mass-loss rate and magnetic topology over the stellar lifetime may alternatively generate obliquity. We hypothesize that the gaseous environments of young, open clusters may provide forcing over sufficient timescales to tilt the astrospheres of young stars, exciting modest obliquities. The more extreme, retrograde stellar obliquities of extrasolar planetary systems likely arise through separate mechanisms.
\end{abstract}
\section{Introduction} 

For centuries, the orbits of the Solar system's planets have been known to occupy a common plane to within 1--2$^\circ$. Such alignment inspired the so-called ``Nebular hypothesis," that planets coalesce from within geometrically-thin disks of gas and dust \citep{Kant1755,Laplace1796}. Images of protoplanetary disks from telescopes such as \textit{Hubble} and \textit{ALMA} have since provided spectacular visual examples of these natal disks \citep{Padgett1999,Andrews2016}. In addition to the planetary coplanarity, if the Sun inherited its angular momentum from the same material as the planets, one would expect the Sun's spin axis to closely coincide with that of the mean angular momentum vector of the planetary orbits--the invariable plane. 

A measurement of the Sun's axial tilt, or obliquity, was unavailable in the days of Kant and Laplace \citep{Carrington1863}, but more recent measurements reveal that the Solar obliquity is 6 degrees \citep{Beck2005,Souami2012,Bailey2016,Gomes2016}\footnote{7 degrees relative to the ecliptic, but 6 relative to the invariable plane.}. This tilt is small when compared to the maximum conceivable range of 0-180 degrees, but is significantly larger than the typical spread of orbital inclinations observed between the orbits of planets--both in our Solar system and in others \citep{Fabrycky2014}. As an illustration, if orbital inclinations follow a normal distribution of standard deviation 2\,degrees, the probability of one member exceeding 6\,degrees by chance alone lies at 0.2\%. Accordingly, an explanation for the origin of the Solar obliquity is required. 

Misalignments between planetary orbits and the spin-axis of their host star, well in excess of the Solar obliquity, have more recently been detected in extrasolar systems \citep{Winn2010,Albrecht2012,Winn2017}. Whereas early attempts to explain the Solar obliquity exist \citep{Tremaine1990,Tremaine1991,Heller1993}, the existence of extrasolar spin-orbit misalignments has prompted a rejuvenation of research  into the factors that influence relative star-orbit orientations. 

Typically, stellar obliquities have been hypothesized to originate from astrophysical mechanisms that tilt the planetary orbits, or their natal disk, as opposed to tilting the star itself. For example, turbulent infall from a molecular cloud core causes the circumstellar disk to undergo a ``wobble" in its orientation, in contrast to the central star that inherits the average angular momentum of the whole stellar core \citep{Bate2010,Fielding2015}. This mechanism is attractive in that it occurs universally across all planet-forming systems, but owing to the high mass of young disks, and strong quadrupole moment of young stars, the star and disk are likely to remain coupled throughout the process \citep{Spalding2014b}, or realign during the subsequent pre-main sequence (PMS) evolution \citep{Spalding2015}.  

A separate scenario calls for a once-present stellar companion, or passing star \citep{Heller1993}, to torque the protoplanetary nebula, thereby tilting the invariable plane relative to the Solar equator \citep{Tremaine1991,Batygin2012,Spalding2014a}. This idea is promising, but would have been more likely to generate larger inclinations than the 6\,$^\circ$ exhibited by the Sun \citep{Batygin2013,Crida2014,Spalding2015}. 

The dynamical role of a stellar companion may equivalently be filled by the gravitational perturbation of a massive, distant perturbing planet well beyond Neptune's orbit \citep{Bailey2016,Gomes2016}. This hypothesis has become more promising in light of growing evidence in support of the existence of such a planet \citep{Batygin2016,Brown2016,Millholland2017,Becker2017}, however, the most recent estimates of the planet's orbit are insufficient to explain the Solar obliquity \citep{Batygin2019}.

In concert with drivers of planetary orbital inclinations, stellar obliquity may arise by way of the direct tilting of the star itself. Mechanisms driving the reorientation of the star are more limited than those affecting planetary orbits. However, during the disk-hosting stage, magnetic star-disk interactions are likely to communicate significant torques between the star and disk, though it remains unclear whether these torques encourage alignment or mis-alignment of the star \citep{Lai2011,Lai2014,Spalding2015}. Accordingly, more work is required to delineate their role, if any, in setting the resultant stellar obliquity.

In this work, we propose a novel mechanism, that the stellar wind, emanating from the star itself, drives a reorientation of the star. The Solar wind is a stream of ionized material emanating from the Sun \citep{Lamers1999}. It is driven by turbulent processes in the corona \citep{Cranmer2005,Cranmer2015}, which involve the interaction between the plasma, the stellar magnetosphere, and the pressure differential driven by the interstellar medium \citep{Parker1958}. 

The earliest models that successfully reproduced the general properties of the Solar wind arose from a one-dimensional, spherically symmetric framework \citep{Parker1958,Parker1965}, with subsequent progression to simple 2-dimensional simulations \citep{Weber1967,Kawaler1988}. However, even at the level of analytic calculations, it became apparent that the wind carries a substantial amount of angular momentum in the form of magnetic stresses, which accelerate the plasma azimuthally \citep{Lovelace2008}.  


As a consequence of the angular momentum carried by the plasma, the stellar wind exerts a braking torque upon the Sun \citep{Kraft1967}, acting as if the wind was co-rotating with the Sun out to a radius known as the Alfv\'en radius, which today lies somewhere between 0.01 and 0.1\,AU\citep{Lovelace2008,Matt2008,DeForest2014}. The magnetic augmentation of angular momentum loss allows the wind alone to drive an order-of-magnitude drop in the rotation rate of Sun-like stars over their main sequence lifetime, despite a relatively small influence upon their masses \citep{Weber1967,Belcher1976,Bouvier2013}. Models of the spin down of main sequence stars have advanced greatly, including 3-dimensional magnetohydrodynamic simulations \citep{Cohen2014,Garraffo2016,Cohen2017,Fionnagain2018,Pognan2018}. However, computational limitations prohibit such simulations from spanning million-year timescales, as is essential in modelling the spin evolution of magnetic stars. 

Though the tendency toward spin-down is not disputed, the time-evolution of spin-rate in Sun-like stars remains an area of active research (see, e.g., \citealt{Metcalfe2019}). If there existed a conserved relationship between age and spin-rate among stars of a specific class, the ages of stars may be determined via gyrochronology. An approximate relationship, known as the Skumanich relation, appears to hold for stars older than 100\,Myr (and younger than 2.5\,Gyr; \citealt{vanSaders2016}) in general \citep{Skumanich1972,Barnes2007,Mamajek2008,Angus2015}, whereby by the spin rate drops as the square root of time. However, this law likely breaks down early in a star's lifetime, when the bulk of angular momentum loss is likely to occur \citep{Garraffo2016,Garraffo2018}.

Of relevance to this work, in modelling efforts that attempt to explain stellar spin-down, the stellar orientation is usually assumed fixed, and only the spin-aligned component of angular momentum is typically investigated, preventing models from revealing tilting torques. In reality, stellar winds are highly variable and turbulent \citep{Tu1995,Cranmer2015,Kiyani2015}, particularly in the young stages \citep{Nicholson2016,Folsom2016}. Turbulence and random perturbations will lead to some degree of stochastic variability in the mean angular momentum axis of the wind \citep{Scherer2001}. Any component of the wind's angular momentum that is directed perpendicular to the star's spin axis will generate a tilting torque, altering the stellar orientation.

Following a qualitative argument, consider that the Sun currently spins about 10 times slower than it once did \citep{Bouvier2013,Fossat2017}. Accordingly, its wind has cumulatively carried away 10 times the Sun's current angular momentum. Only 1\%, on average, of that wind-driven torque needs to have acted perpendicular to the spin axis over the past 4.5\,Gyr for the Sun's current orientation to differ by several degrees from its primordial orientation.

Some evidence exists that the angular momentum of the Solar wind is, at least instantaneously, not perfectly aligned along the spin axis \citep{Scherer2001,Smith2001}. Thus, even if the star-formation process resulted in a perfectly-aligned system of planetary orbits around an aligned stellar equator, stochasticity in the stellar wind may in principle drive a deviation from this aligned architecture over its subsequent evolution. 

In this paper, we develop a simple model of the evolution of a star's spin axis under the action of stochastically-varying torques owing to the stellar wind. We allow these torques to deviate from perfectly aligned with the spin axis by a small angle, of root-mean-squared magnitude $\epsilon_0$, that varies stochastically in direction and magnitude over a typical autocorrelation timescale $\tau_c$. We derive a simple expression relating the values of these two parameters to the probability distribution of resulting stellar obliquities, including the likelihood of Solar-like obliquities. We finish with a discussion of the possibility of favourable parameters to occur in the early evolution of the Solar system, specifically within the stellar birth cluster, where large gas densities may warp the heliosphere coherently, and drive a long-term tilting torque. 

\section{Methods}

In this section, we describe a simple model where the angular momentum of a star changes in response to torques originating from the stellar wind. Much previous work has computed the spin-down torque exerted by the stellar wind's interaction with the stellar magnetosphere \citep{Weber1967,Belcher1976,Lamers1999,Matt2008,Garraffo2016}. For a star with mass-loss rate $\dot{M}_\star$ and spin rate $\Omega_\star$, the rate at which angular momentum is lost (denoted $\dot{L}$) takes the form
\begin{align}
\dot{L}=\mu \dot{M}_\star r_A^2 \Omega_\star,
\end{align} 
where $\mu$ is a constant of order unity and $r_A$ is the Alfv\'en radius, defined as the point where the stellar wind radial velocity reaches the Alfv\'en velocity. The magnitude of $\mu$ varies depending upon the assumed magnetic topology \citep{Garraffo2016} but its value will not be important to our calculations below. 

 In most previous treatments, the torque $\dot{L}$ is assumed to act perfectly anti-parallel to the stellar spin axis, leading to a deceleration of $\Omega_\star$, but no change in the stellar orientation. However, the stellar wind launching mechanisms are highly turbulent \citep{Cranmer2015}, and the magnetosphere of the Sun itself is time-variable and non-azimuthally symmetric \citep{Hoeksema1982,Scherer2001}. Therefore, it is unlikely that the angular momentum lost is directed identically along the rotation axis. 
 
\begin{figure}[ht!]
\centering
\includegraphics[trim=0cm 0cm 0cm 0cm, clip=true,width=1\columnwidth]{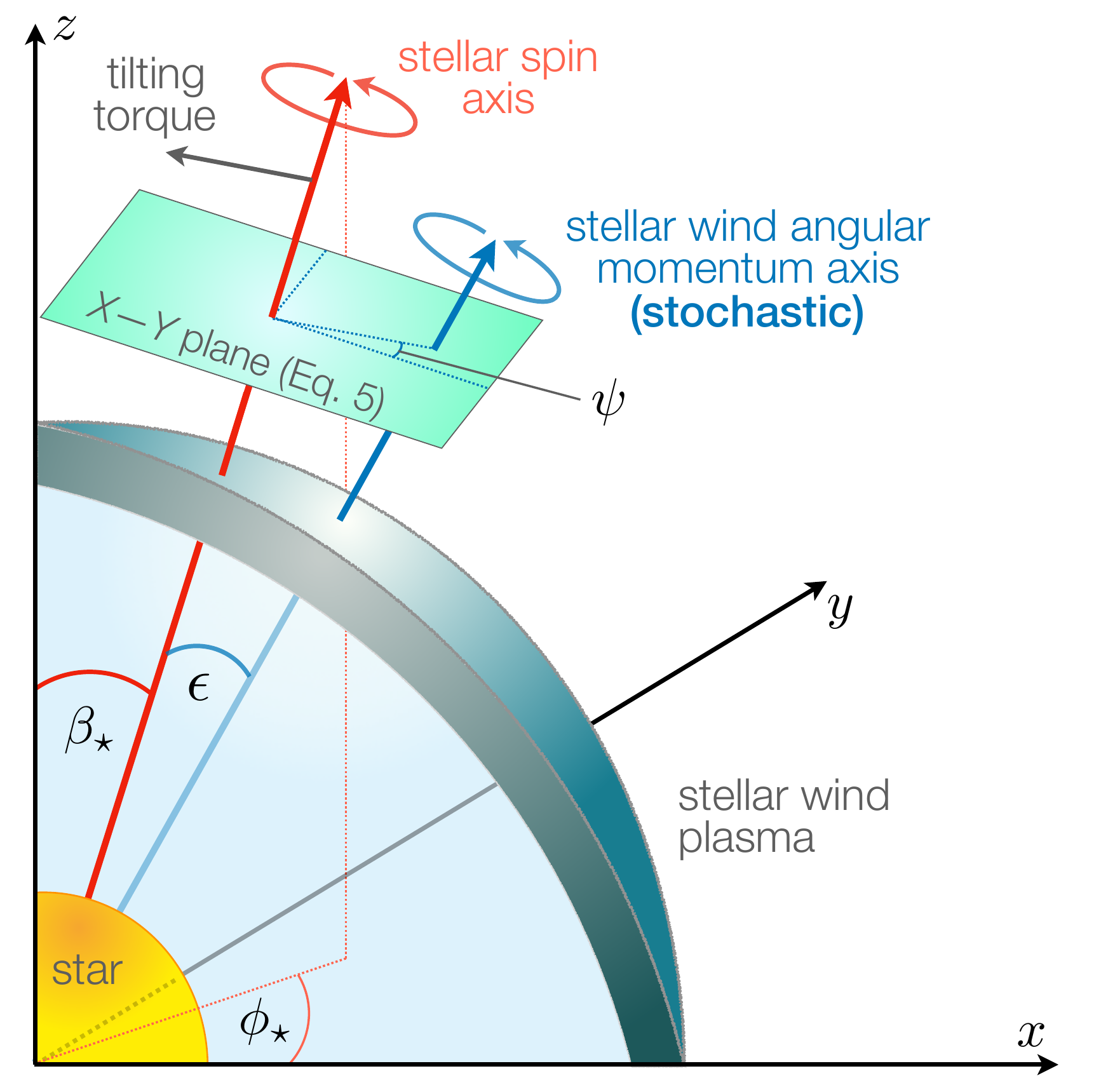}
\caption{A schematic of the modelled scenario. The star spins about the red axis, which is inclined by $\beta_\star$ to the $z-$axis. Its projection into the $x-y$ plane makes an angle $\phi_\star$ with the $x-$axis. The star is emitted a wind (blue shell) that possesses a mean angular momentum axis given by the long blue arrow. We construct an orthogonal coordinate system defined by an $X-Y$ plane (green rectangle) that is place with the stellar spin axis as the normal vector. Relative to this local coordinate system, the stellar wind's mean spin is offset by an angle $\epsilon$, with a projection into the $X-Y$ plane that makes an angle $\psi$ with the $X-$axis.}
\label{f1}
\end{figure}

 In our modelling below, we consider the consequences of a slightly off-axis angular momentum flux. Specifically, we assume that the magnitude of the torque is $\dot{L}$ as defined above, but its direction makes an angle $\epsilon$ with the stellar spin axis, and is directed at an angle $\psi$ when projected onto a plane normal to the stellar spin axis (see Figure~\ref{f1} for a schematic). Both $\epsilon$ and $\psi$ vary stochastically, as an Ornstein-Uhlenbeck process with autocorrelation time $\tau_c$, reflecting a random process that nevertheless exhibits some degree of ``memory" over timescales of $\tau_c$ \citep{Gardiner2009} \footnote{The increased power at longer timescales gives this type of forcing the name ``red noise". In this context, white noise may be thought of as Ornstein-Uhlenbeck noise with $\tau_c\rightarrow0$, i.e., memoryless.}. The root-mean-square value of $\epsilon$ (or its standard deviation) is set equal to $\epsilon_0$ whereas $\psi$ is isotropically-varying.

\subsection{Spin-axis evolution}

Consider a spherical coordinate system where the stellar angular momentum vector (assumed parallel to the angular velocity vector) takes the form
\begin{align}
\mathbf{J}&=k M_\star \Omega_\star R_\star^2\,\mathbf{\hat{e}}_3\nonumber\\
&\equiv J\,\mathbf{\hat{e}}_3
\end{align}
Above, $\mathbf{\hat{e}}_3$ is a unit vector that lies parallel to the spin axis, and points outward from the origin (see Figure~\ref{f1}). The star's mass and radius are, respectively $M_\star$ and $R_\star$, while the dimensionless moment of inertia parameter $k$ reflects the star's inner structure. The projection of the spin vector makes an angle $\phi_\star$ with the $x-$axis and makes an angle $\beta_\star$ to the $z-$axis.

It is simplest to work in the body-centric frame, i.e., to consider the evolution of the spin axis in terms of spherical unit vectors $\mathbf{\hat{e}}_\theta$ and $\mathbf{\hat{e}}_\phi$, such that torques directed along the former alter obliquity $\beta_\star$ and torques directed along the latter alter $\phi_\star$. Using vector relationships, one can show that
\begin{align}\label{three}
\frac{d\mathbf{J}}{dt}&=\dot{J}\mathbf{\hat{e}}_3+\mathbf{\omega}\times\mathbf{\hat{e}}_3\nonumber\\
&=\dot{J}\mathbf{\hat{e}}_3+J \dot{\beta}_\star \mathbf{\hat{e}}_\theta+J \dot{\phi}_\star \mathbf{\hat{e}}_\phi,
\end{align}
where $\mathbf{\omega}$ is a vector with components $\dot{\beta}_\star$ in the $\theta-$direction and $\dot{\phi}_\star$ in the $\phi-$direction\footnote{These are otherwise referred to as Euler angles \citep{Euler1776,Goldstein2011}}.
As mentioned above, we assume that the spin-down torque, $\dot{L}$ is directed off-axis by an angle $\epsilon$. The direction of the torque is expressed in terms of an angle $\psi$, which is found by projecting the torque onto a plane that is normal to the stellar spin axis. Expressed as a vector (denoted $\mathbf{M}$) this torque takes the form
\begin{align}\label{four}
\mathbf{M}=-\dot{L}\big(\mathbf{\hat{e}}_3 \cos(\epsilon)+\mathbf{\hat{e}}_\theta \sin\epsilon\cos\psi+\mathbf{\hat{e}}_\phi\sin\epsilon\sin\psi\big).
\end{align}
Equating expressions~\ref{three} and~\ref{four} yields the evolutionary equations
\begin{align}
\frac{d}{dt}\bigg[k M_\star \Omega_\star R_\star^2\bigg]&=-\mu \dot{M}_\star r_A^2\Omega_\star \cos(\epsilon)\nonumber\\
\frac{d\beta_\star}{dt}&=-\bigg[\frac{\mu \dot{M}_\star r_A^2}{k M_\star R_\star^2}\bigg]\sin(\epsilon)\cos(\psi)
\end{align}
It is convenient to make the assumption that the offset angle $\epsilon\ll1$, and to define new stochastic variables, which are the projections of the tangential torque when $\epsilon\ll1$, as  
\begin{align}
X_t\equiv \epsilon(t)\cos{\psi(t)}\,\,\,\,\,\,\,\,Y_t\equiv \epsilon(t)\sin{\psi(t)}.
\end{align}

Next, we make the assumption that $kM_\star R_\star^2$ is time-independent throughout this time, facilitating the simplifying approximation that $\dot{J}=kM_\star R_\star^2\dot{\Omega}_\star$. Accordingly, we may define a time-dependent, characteristic spin-down timescale
\begin{align}\label{tausp}
\tau_{sp}\equiv\bigg|\frac{\Omega_\star}{\dot{\Omega}_\star}\bigg|=\frac{k M_\star R_\star^2}{\mu \dot{M}_\star r_A^2}.
\end{align}
It is important to note that expression~\ref{tausp} for $\tau_{sp}$ exhibits numerous forms of time-dependence that are not understood in detail. In particular, the mass-loss rate, Alfv\'en radius and prefactor $\mu$ all change with time. Given our incomplete knowledge of many of the underlying mechanisms driving stellar spin-down \citep{Kraft1967,Bouvier2013,Cohen2017,Garraffo2018}, we shall utilize empirical observations of the evolution of stellar spin rates with time in order to parameterize $\tau_{sp}$.
 
After the above substitutions, the evolution of the stellar spin is fully described by the 3 stochastic differential equations
\begin{align}\label{generalBeta}
\frac{d\beta_\star}{dt}&=-\frac{X_t}{\tau_{sp}}\nonumber\\
\frac{d\phi_\star}{dt}&=-\frac{Y_t}{\tau_{sp}}\nonumber\\
\frac{d\Omega_\star}{dt}&=-\frac{\Omega_\star}{\tau_{sp}}.
\end{align}

Our goal is to model the star's response to fluctuations in $\epsilon$ that vary over a typical timescale $\tau_c$. We describe this process by prescribing $X_t$ and $Y_t$ to evolve according to independent Ornstein-Uhlenbeck stochastic processes, with individual autocorrelation times $\tau_c$ and variances set to generate $\langle X_t^2+Y_t^2\rangle=\epsilon_0^2$ at times $t\gg \tau_c$. These statistical parameters are consistent with the following stochastic differential equations (see e.g. \citealt{Gardiner2009}),
\begin{align}\label{WindVar}
dX_t=-\frac{X_t}{\tau_c}dt+\sqrt{\frac{1}{\tau_c}}\epsilon_0 dW_t\nonumber\\
dY_t=-\frac{Y_t}{\tau_c}dt+\sqrt{\frac{1}{\tau_c}}\epsilon_0 dW_t,
\end{align}
where $W_t$ is a normally-distribution, zero-mean Weiner process with unit variance. 

\subsection{Magnetic braking}

In order to solve equations~\ref{generalBeta} and~\ref{WindVar}, it is necessary to parameterize the time-evolution of $\tau_{sp}$. The general pattern of stellar rotational evolution may be described as a short ($\sim1-10$\,Myr) disk-hosting phase, during which the spin rate does not appear to evolve significantly \citep{Bouvier2013}. This is typically considered to be a result of ``disk-locking" \citep{Shu1987}, whereby magnetic interactions with the protoplanetary disk maintain a constant stellar spin rate against contraction driven spin-up. Some evidence in support of disk-locking lies in a slight dichotomy between disk-lacking young stars spinning somewhat faster ($\sim1-7$\,days) than disk-hosting stars ($\sim3-10$\,days; see \citealt{Bouvier2014} and references therein). Physically, if the disk dissipates before the star has completed its contraction, as expected for the Sun, then the subsequent few million years are thought to lead to contraction-driven spin-up \citep{Pognan2018}.

Once contraction ceases, and the disk has dissipated, the star's spin evolution is dominated by magnetic braking. In our modeling, we consider the evolution of the star's spin axis beginning from this point, and consider a range of initial spin rates between $1-10$\,days. From the functional form of equation~\ref{generalBeta}, with $\tau_{sp}$ in the denominator, tilt occurs more rapidly when the spin-down time is short. Heuristically, if we suppose that the Sun began spinning at 1\,day, it will typically exhibit a shorter initial value of the spin down time, than if it began spinning at 10\,days, given the constraint that its current spin period must equal $24.5$ (at the equator). Owing to this constraint, and the inverse dependence of $\dot{\beta}_\star$ upon $\tau_{sp}$, our solutions below shall conclude that the faster the initial spin of the Sun, the more tilting is expected to occur.

\subsection{Time-evolution of stellar spin}

Among stars older than several 100 million years, the so-called Skumanich relation broadly applies \citep{Skumanich1972}, whereby the spin rate follows the time evolution 
\begin{align}
\Omega_\star(t)= \Omega_{\star,i} \bigg(\frac{t}{t_i}\bigg)^{-\gamma},
\end{align}
with $\gamma=1/2$ \citep{Skumanich1972,Cohen2014}. The parameters $t_i$ and $\Omega_{\star,i}$ are the time and spin-rate at a standard reference point. 

The Skumanich relationship has generally held up to more recent investigations \citep{Barnes2007,Mamajek2008,Angus2015}. Nevertheless, young stars appear to exhibit a wide distribution of rotation periods \citep{Bouvier2013} that cannot be described by a single dependence upon time. Even among older stars there remains a subset that appear to rotate faster than predicted by this law \citep{Garraffo2016,Garraffo2018}. Moreover, difficulties in assigning ages to stars hinder a thorough test of the details of spin-down, despite an ever-growing set of spin-period observations \citep{McQuillen2014,vanSaders2016,vansaders2019}.

Let us suppose that the appropriate braking law follows the form
\begin{align}
\dot{\Omega}_{\star}= k\Omega^{3},
\end{align}
where $k$ is a complicated function of stellar parameters. In order to make progress, we shall assume that $k$ is constant, and fix its value such as to reproduce the modern-day Sun's spin rate. This will allow us to make analytic progress, but we shall return to the effect of time-dependence in $k$ in the discussion \citep{Garraffo2018}. The solution to this equation, for a star with initial spin rate $\Omega_0$ and time $t_0=0$ is
\begin{align}
\Omega_\star(t)=\Omega_0\big[1-2k t\Omega_0^2\big]^{-1/2},
\end{align}
which reproduces the modern-day solar values $\Omega_0=\Omega_\odot$ and $t_\odot$ with the choice
\begin{align}
k=-\frac{1}{2\Omega_0^2 t_\odot}\bigg[\bigg(\frac{\Omega_0}{\Omega_\odot}\bigg)^2-1\bigg],
\end{align}
and so
\begin{align}
\Omega_\star(t)=\Omega_0\Bigg[1+ \frac{t}{t_\odot}\bigg(\bigg[\frac{\Omega_0}{\Omega_\odot}\bigg]^2-1\bigg)\Bigg]^{-1/2}.
\end{align}
Notice that the form above reproduces the Skumanich relation for constant $k$ as $t\rightarrow\infty$.

We may now compute the spin-down timescale from the definition~\ref{tausp}, which takes the form
\begin{align}\label{tauSpin}
\tau_{sp}(t)=2t+2t_\odot\bigg[\frac{P_0^2}{P_\odot^2-P_0^2}\bigg],
\end{align}
where we have introduced the spin periods of the modern Sun $P_\odot=24.5$\,days (at the equatorial surface, \citealt{Thompson2003}) and the initial, young Sun $P_0$. 
As $t\rightarrow 0$, we find that the initial spin-down time approaches
\begin{align}\label{InitialTau}
\tau_{sp,0}&\approx 2t_\odot \Bigg[\frac{P_0}{P_\odot}\Bigg]^2\nonumber\\
&\approx 14\bigg(\frac{P_0}{\textrm{day}}\bigg)^2\textrm{Myr},
\end{align}
and monotonically increases from there. Accordingly, as noted qualitatively above, the minimum spin down timescale experienced by the star occurs at the initial time, and is determined entirely by the period assumed at $t=0$.

Now that we possess a simple expression for the spin-down timescale, we may return to the evolutionary equation of the spin-axis, and develop a semi-analytic solution for the tilt angle $\beta_\star$, or rather its root-mean square (rms) value.

\subsection{Approximate analytic solution}

Utilizing the analytic form of the spin-down time, we now proceed to compute a statistical description of the stellar tilt axis evolution. To begin, we compute the change in stellar spin axis $\delta\beta$ expected over a single correlation time $\tau_c$. Throughout this interval, we assume that $X_t=\epsilon_0$; provided we only integrate over a correlation time, we do not expect the value of $X_t$ to vary substantially, such that
\begin{align}
\frac{d\beta_\star(t)}{dt}\approx-\frac{\epsilon_0}{\tau_{sp}(t)}.
\end{align}
Maintaining this configuration over a timescale $\tau_c$, the stellar obliquity will change by an amount $\delta \beta$, given by
\begin{align}
\delta\beta(t)&\approx \int_{t}^{t+\tau_c}\frac{\epsilon_0}{\tau_{sp}(t')}dt'\nonumber\\
&=\frac{\epsilon_0}{2}\ln\Bigg[1+\frac{\tau_c}{t+t_\odot Q^2}\Bigg],
\end{align}
where we have defined the ratio
\begin{align}
Q=\frac{P_0}{\sqrt{P_\odot^2-P_0^2}}\approx \frac{P_0}{P_\odot}.
\end{align}

Physically, we may consider this change in orientation as a stochastic ``kick" $\delta\beta$ of the stellar spin axis over a timescale $\delta t=\tau_c$. Summing up many such kicks, the process may be modelled as a random walk with a time-dependent diffusion coefficient \citep{Chandrasekhar1943,Pettibone2009}
\begin{align}
D(t)\equiv \frac{\delta \beta(t)^2}{\delta t}=\frac{\epsilon_0^2}{4\tau_c}\ln\Bigg[1+\frac{\tau_c}{t+t_\odot Q^2}\Bigg]^2.
\end{align}
 After a time $t$ has passed, we expect the root-mean-square value of $\beta_\star$ (essentially the standard deviation) to grow diffusively as
\begin{align}\label{AnalyticSol}
\langle\beta_\star^2\rangle\big|_t&\approx \int_{0}^{t}D(t')dt'\nonumber\\
\langle\beta_\star^2\rangle\big|_t&=\frac{\epsilon_0^2}{4} \int_{t_\odot Q^2/\tau_c}^{(t+t_\odot Q^2)/\tau_c}\Bigg[\ln\bigg(1+\frac{1}{u}\bigg)\Bigg]^2du,
\end{align}
where we introduce the scaled, dummy variable $u\equiv (t'+t_\odot Q^2)/\tau_c$. The integrand above possesses no closed-form solution (though approximations exist in the regimes where $u\gg1$ or $u\ll1$). 

We are interested in the probability of the modeled process to reproduce the Sun's current obliquity. Thus, we shall consider only the long-term evolution of the spin-axis ($t/\tau_c\rightarrow\infty$). In the long-term limit, the mean-squared value of stellar obliquity depends only upon the dimensionless parameters $t_\odot Q^2/\tau_c=\tau_{sp,0}/(2\tau_c)$ and $\epsilon_0$. For convenience, we define the integral function
\begin{align}\label{transferFunc}
\mathcal{F}(b)\equiv\sqrt{\int_{b/2}^{\infty}\Bigg[\ln\bigg(1+\frac{1}{u}\bigg)\Bigg]^2du},
\end{align}
leading to the solution for the root-mean-squared stellar obliquity, 
\begin{align}\label{BetaSol}
\boxed{\sqrt{\langle\beta_\star^2\rangle}\big|_{t\rightarrow\infty}=\frac{1}{2}\epsilon_0\mathcal{F}\bigg(\frac{\tau_{sp,0}}{\tau_c}\bigg).}
\end{align}

The function $\mathcal{F}$ can be thought of as a scaling coefficient that determines the fraction of the misalignment $\epsilon_0$ that is imprinted upon the stellar obliquity. We plot its value in Figure~\ref{f2}. 

\subsection{Distribution of obliquities}

Having computed an approximate evolution for the variance of stellar obliquities, we may now derive a probability distribution of obliquities expected to result from the stellar wind-driven torques. Let us suppose that the probability of the star attaining a tilt between $\beta_\star$ and $\beta_\star+d\beta_\star$ is described by a Gaussian distribution:
\begin{align}\label{Gauss}
P(\beta_\star)d\beta_\star=\sqrt{\frac{1}{2\pi \sigma^2}}\exp\bigg[-\frac{\beta_\star^2 }{2 \sigma^2}\bigg]d\beta_\star,
\end{align}
where the standard deviation $\sigma$ is well-approximated by the root-mean-square stellar obliquity (Equation~\ref{BetaSol}), 
\begin{align}
\sigma=\sqrt{\langle\beta_\star^2\rangle}\big|_{t\rightarrow\infty}=\frac{\epsilon_0}{2}\mathcal{F}\bigg(2\frac{t_\odot}{\tau_c}Q^2\bigg).
\end{align}
After integrating, we arrive at an expression for the probability that the magnitude of the stellar tilt exceeds $\beta'$:
\begin{align}\label{PDF}
\boxed{
P(\beta_\star>\beta')=1-\textrm{Erf}\Bigg[\frac{\sqrt{2}\beta'}{\epsilon_0\mathcal{F}(2t_\odot Q^2/\tau_c)}\,\Bigg].}
\end{align} 
The expression above cannot be written in closed form, but its value is easily computed for any given parameters. Note that the distribution only depends on the ratio $\tau_{sp,0}/\tau_c$ and $\epsilon_0$ and $\tau_{sp,0}$ depends only upon the initial spin period $P_0$.

We illustrate the probability of the tilt exceeding the Solar value of $\beta'=\beta_\odot\approx6^\circ$ in Figure~\ref{f4}, as a function of the ratio $\tau_{sp,0}/\tau_c$ ($y-$axis), and $\epsilon_0$ ($x-$axis). In general, for Solar-like obliquities to occur in 50\% of cases, offsets of $\epsilon\gtrsim10^\circ$ are required to persist over timescales $\tau_c\gtrsim \tau_{sp,0}$. Later, we discuss how realistic these parameters are.

As we will show using numerical simulations, the above analytic approximations provide a remarkably accurate representation of the full integrations. Thus, we conclude this subsection by highlighting that the expected distribution of stellar obliquity, arising from a stochastically-fluctuating stellar wind, is fully described to a close approximation by Equation~\ref{PDF}. 

\begin{figure}[ht!]
\centering
\includegraphics[trim=0cm 0cm 0cm 0cm, clip=true,width=1\columnwidth]{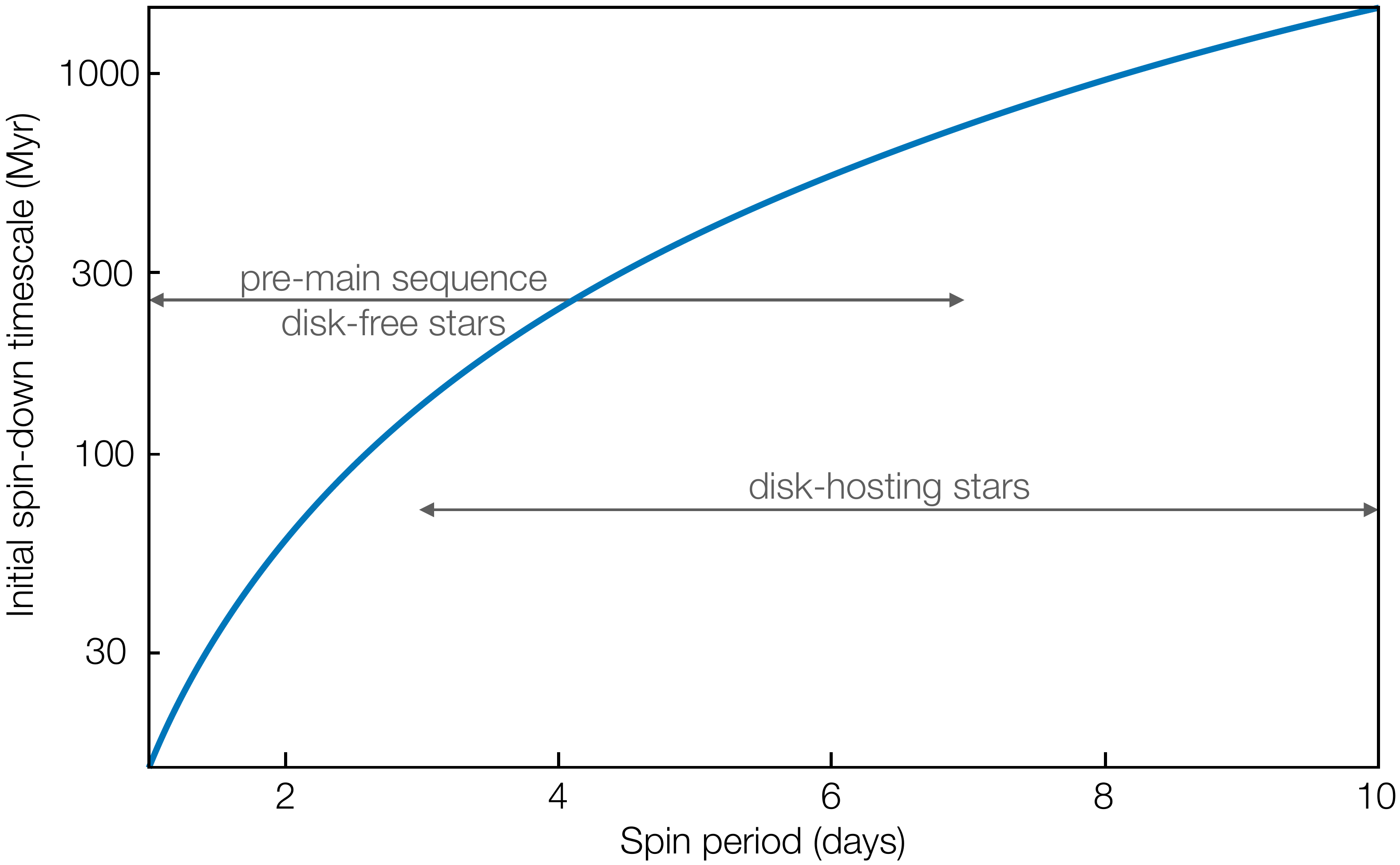}
\caption{The relationship between the initial spin-down timescale ($\tau_{sp}$ evaluated at $t=0$) as a function of the initial spin period of the star (as defined in equation~\ref{tauSpin}). On the figure, arrows indicate the typical range of initial spin-periods for young stars both with and without disks \citep{Bouvier2014}. The faster spins of disk-less stars suggest that stars tend to begin their magnetic braking evolution from the lower end of the observed spin periods indicated, and so begin with spin-down timescales ranging from 10-100\,Myr.}
\label{f2_new}
\end{figure}

\begin{figure}[ht!]
\centering
\includegraphics[trim=0cm 0cm 0cm 0cm, clip=true,width=1\columnwidth]{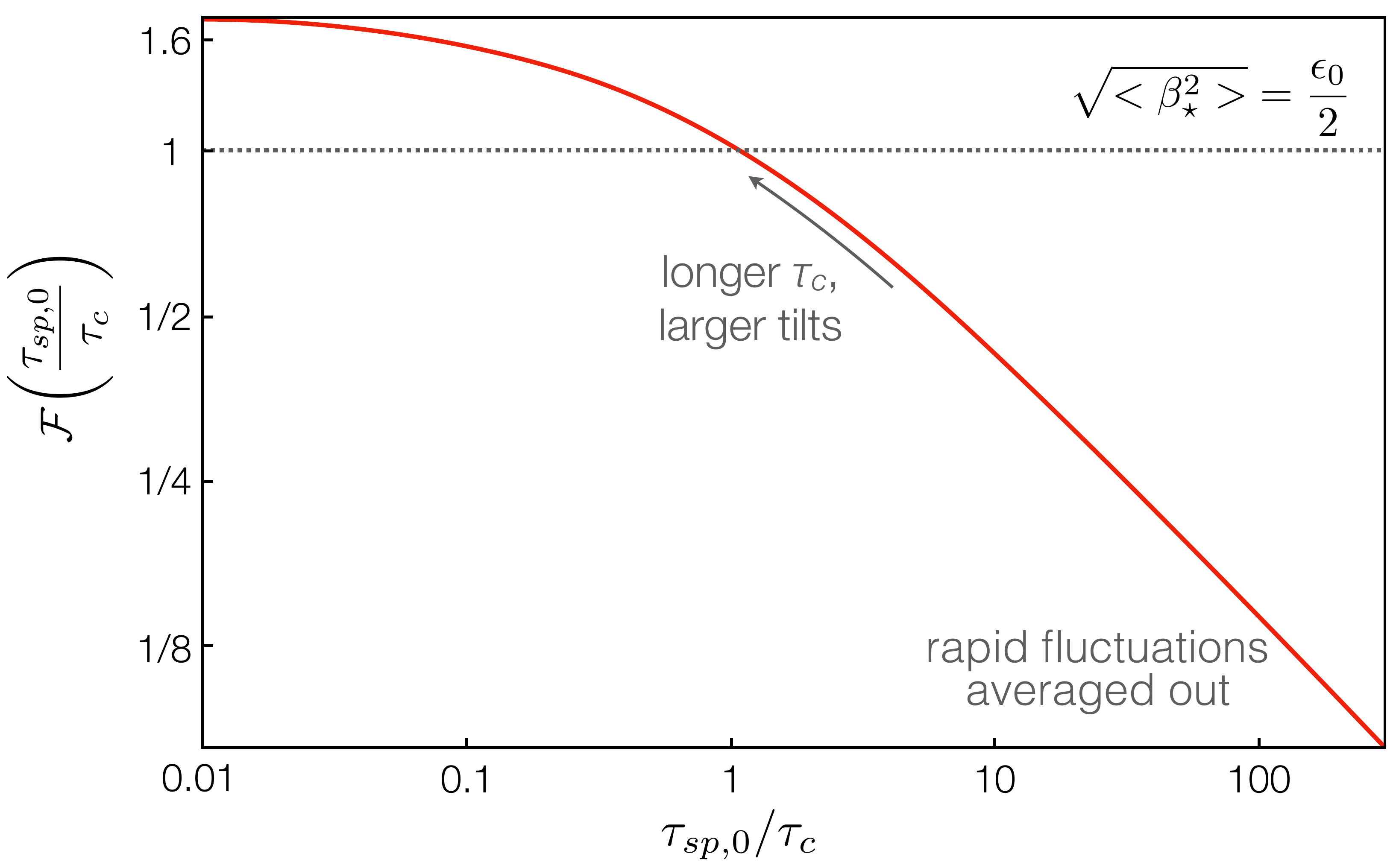}
\caption{The value of $\mathcal{F}$ in Equation~\ref{transferFunc} as a function of the ratio $r=\tau_{sp,0}/\tau_c$. Where $\mathcal{F}$ equals unity translates to a root-mean-square stellar obliquity of half the root-mean-square offset angle $\epsilon_0$. Only for small values of $\tau_{sp,0}$, or long correlation time perturbations $\tau_c$, are significant obliquities typically excited.}
\label{f2}
\end{figure}

\begin{figure}[ht!]
\centering
\includegraphics[trim=0cm 0cm 0cm 0cm, clip=true,width=1\columnwidth]{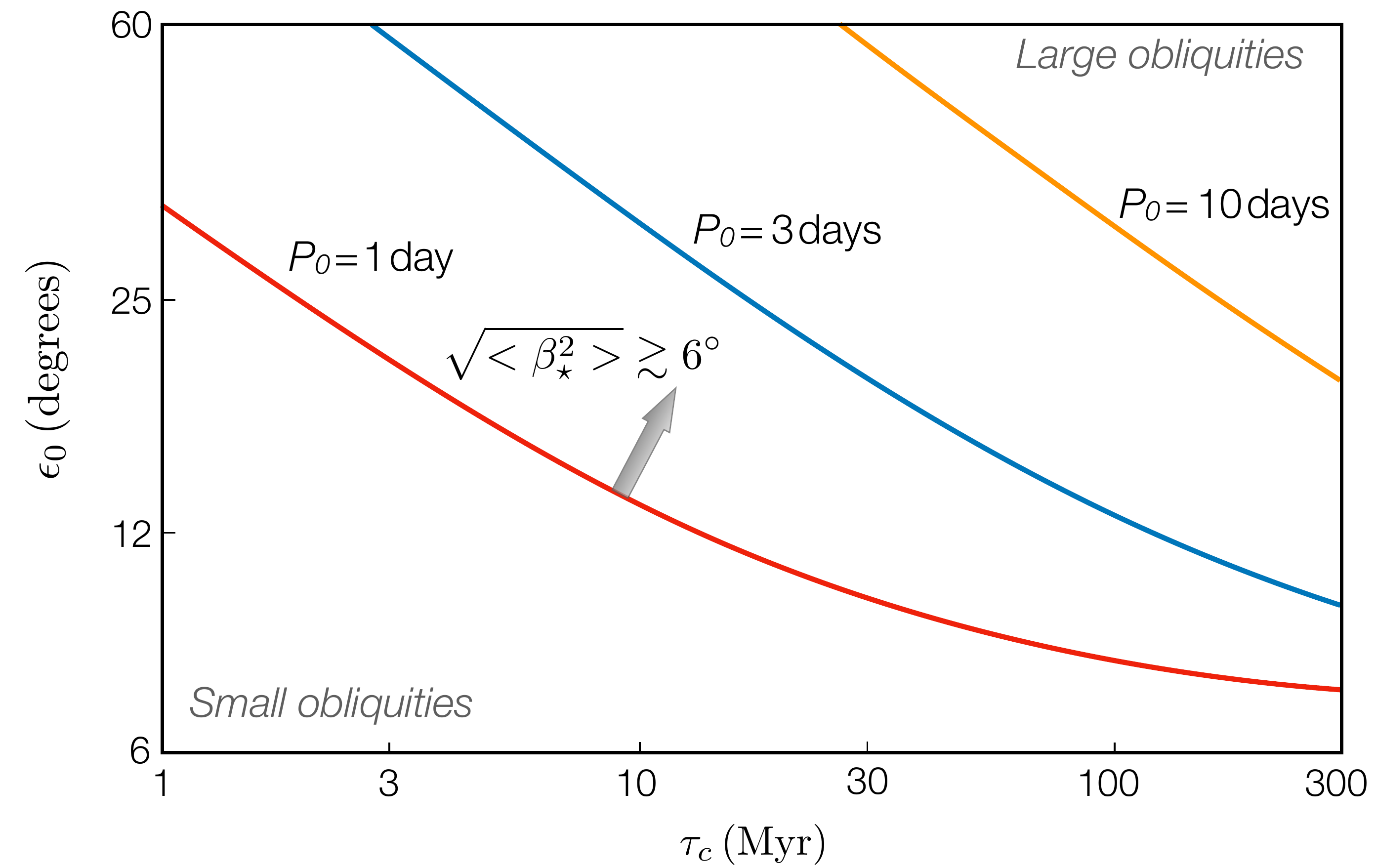}
\caption{The required root-mean-squared offset between the angular momentum vector of the solar wind and the stellar spin axis in order to generate root-mean-square stellar obliquities greater than that of the Sun ($6^\circ$). The lines denote different choices for the initial stellar rotation period ($P_0=1$\,day, red; $P_0=3$\,days, Blue; $P_0=10$\,days, orange). These 3 periods span the likely range of stars throughout pre-main sequence evolution\citep{Maeder2000,Bouvier2014}.}
\label{f3}
\end{figure}

\begin{figure}[ht!]
\centering
\includegraphics[trim=0cm 0cm 0cm 0cm, clip=true,width=1\columnwidth]{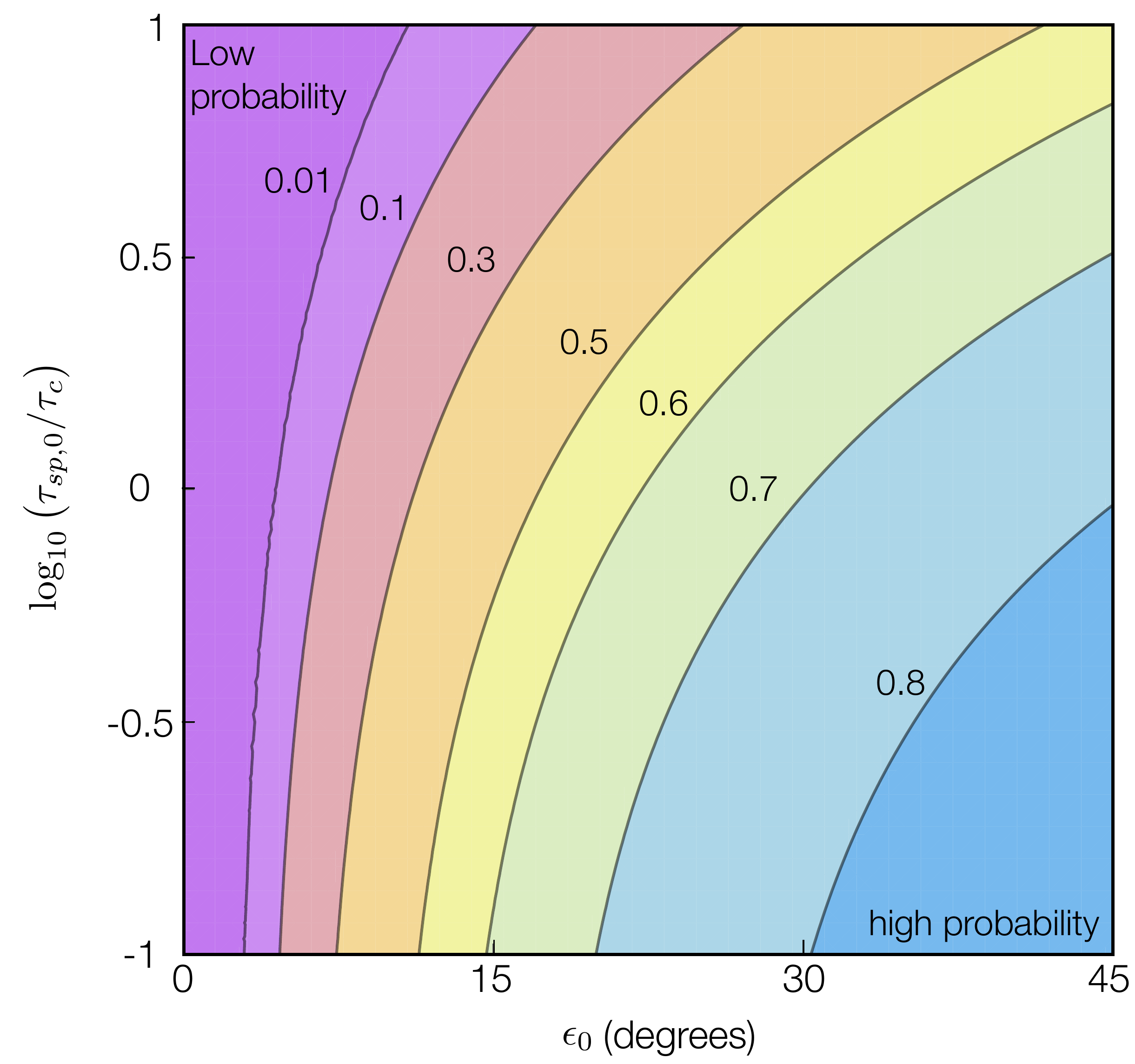}
\caption{Contours showing the probability that the stellar obliquity reaches or exceeds $6^\circ$ as a function of the stellar wind offset $\epsilon_0$ and the ratio of initial spin-down timescale to the wind's turbulent correlation time. If the Solar obliquity is to be ascribed to stellar wind-tilting to any reasonable likelihood, the correlation time must exceed $\tau_{sp,0}$. Additionally, the typical offset $\epsilon_0$ must be of order $10^\circ$ or larger.}
\label{f4}
\end{figure}


\begin{figure}[ht!]
\centering
\includegraphics[trim=0cm 0cm 0cm 0cm, clip=true,width=1\columnwidth]{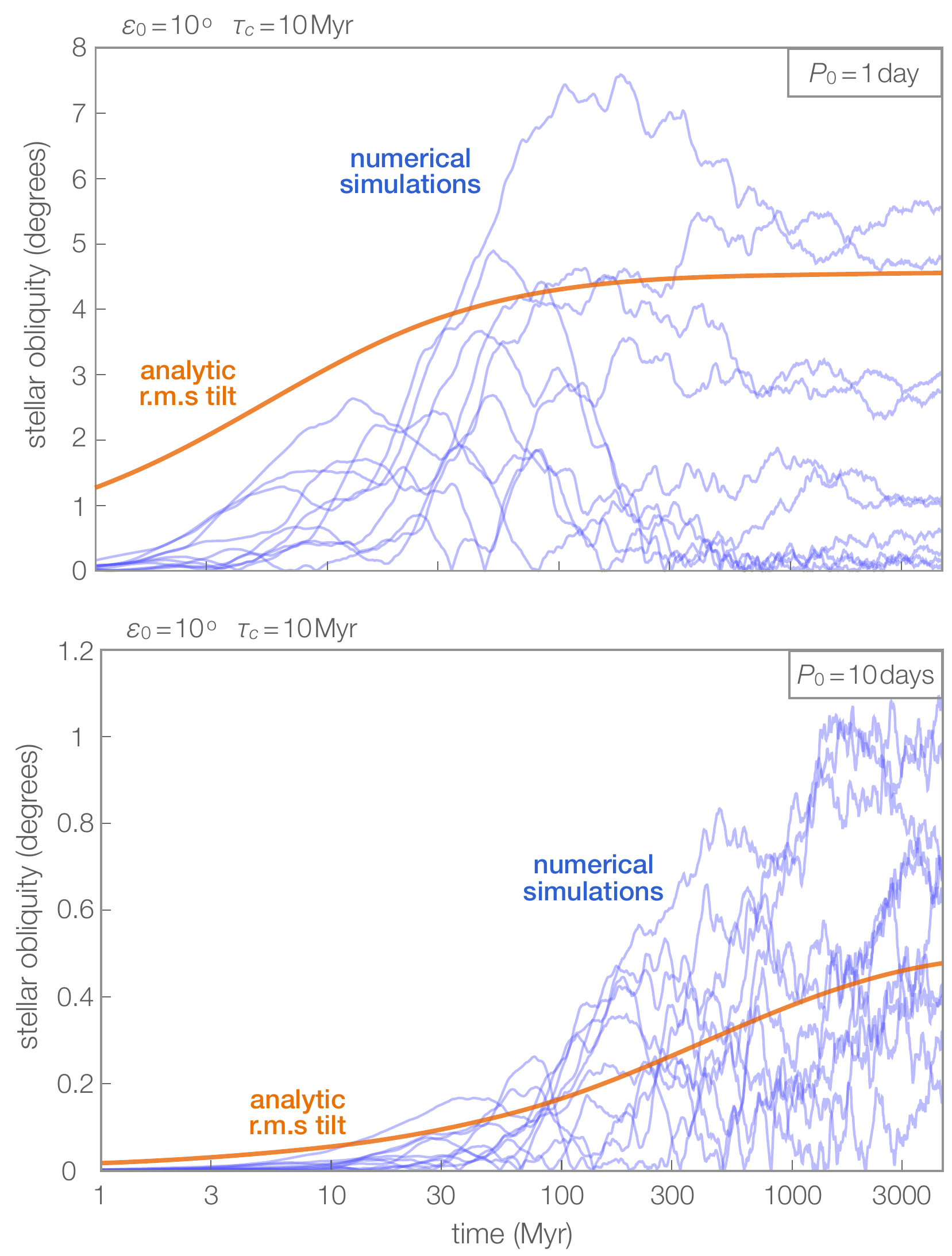}
\caption{Blue trajectories are the results from numerical simulations of the evolution of the star's spin axis in response to a stochastically-varying wind direction. The red line is the approximate solution for the root-mean-square obliquity as obtained from Equation~\ref{BetaSol}. We set $\epsilon_0=10^\circ$ and $\tau_c=10$\,Myr, but choose $P_0=1\,$day for the upper panel and $P_0=10\,$days for the lower panel. Only the first 10 simulations of 100 are shown in each case, and note the change in vertical axis scales. Solar-like obliquities arise only with the faster initial spin period of 1\,day.}
\label{f5}
\end{figure}

\subsection{Numerical simulation}

The computation above derived an approximate expression for the required statistical properties of the stellar wind, in order to reproduce the observed stellar obliquity. In this section, we perform numerical simulations of the stochastic differential Equations~\ref{generalBeta} and~\ref{WindVar}. This allows us to analyze the resulting distribution of stellar obliquities, and validate the simple calculations above. 

\begin{figure}[ht!]
\centering
\includegraphics[trim=0cm 0cm 0cm 0cm, clip=true,width=1\columnwidth]{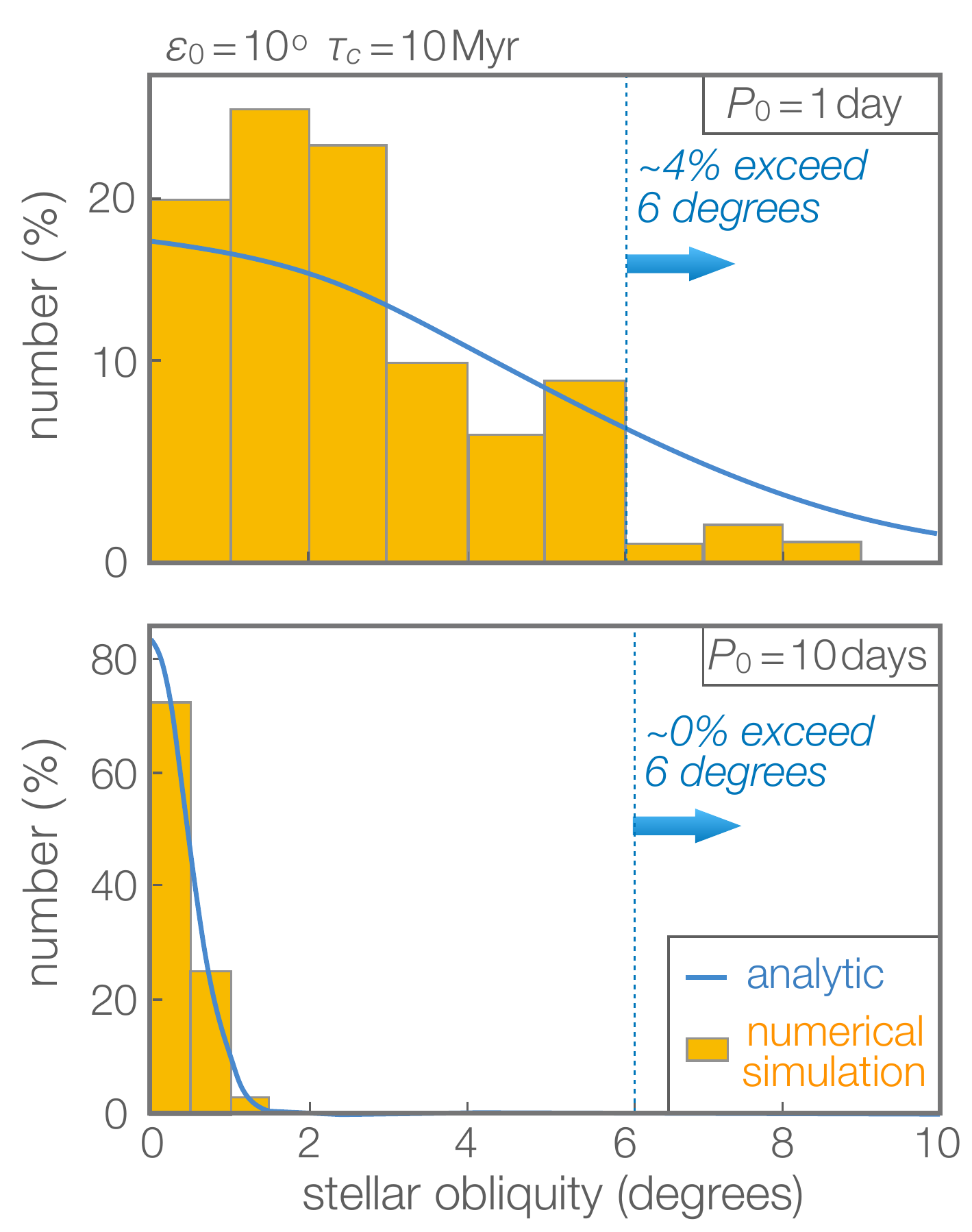}
\caption{Yellow histograms indicate the probability density function of stellar obliquities within the 100 trajectories numerically integrated for 4.5\,Gyr. We set $\epsilon_0=10^\circ$ and $\tau_c=10$\,Myr, and choose $P_0=1\,$day for the upper case and $P_0=10\,$days for the lower case. The blue curve indicates an approximate Gaussian distribution with width determined from our approximate solution~\ref{Gauss}. The numerical simulations agree well with our approximate calculations.}
\label{f6}
\end{figure}

We performed 2 stochastic simulations, each lasting 4.5\,Gyr, but with one beginning at a rotation period $P_0=1\,$day and the other from $P_0=10$\,days. From then, the evolution is computed according to the spin-down time evolution given in relationship~\ref{tauSpin}. We split into about $2\times10^4$ time-steps and integrated using a simple Euler scheme. The parameters $\epsilon_0=10^\circ$ and $\tau_c=10\,$Myr were maintained throughout. For each simulation, 100 trajectories were obtained, of which the first 10 are illustrated in Figure~\ref{f5}. Over-plotted on these trajectories is the analytic solution derived above (Equation~\ref{BetaSol}) for the root-mean-squared stellar obliquity, appears to describe well the median of the trajectories.

After having obtained 100 trajectories for both sets of parameters, we generated a histogram of the resulting stellar obliquities for each case, illustrated in Figure~\ref{f6}. We compare these numerically-generated histograms to the Gaussian approximation derived above (Equation~\ref{Gauss}), indicating the approximate agreement between the expected and true distributions. Furthermore, these results reinforce the requirement that Solar-like obliquities occur frequently only when $\tau_c>t_{sp,0}$. A spin-period of 1\,day corresponds to $\tau_{sp,0}=14\,$Myr, which is slightly in excess of $\tau_c$. Thus 6$^\circ$ obliquities only occur in about 4\% of cases, but almost never with 10\,day initial spin-periods.


It was clear from the differential equation alone that larger obliquities are excited when the spin-down time is short. However, the solution above indicates that the growth rate of the rms obliquity is fastest for times shorter than the correlation time. Combined, larger obliquities arise from a larger ratio of $\tau_{c}/\tau_{sp}$.

We assumed a Skumanich-like braking law right back to $t=0$, however numerous mechanisms during early stellar evolution may cause significant deviations from such a relationship. In particular, it has been proposed \citep{Garraffo2016,Garraffo2018} that for the most rapidly-rotating initial states, the magnetic field topology may be altered such as to include higher-order components than a simple dipole. This weakens the braking effect, early-on, allow spin-up to shorter periods. In order to arrive at a Solar-like spin rate, such a star must later go through a period of rapid spin-down. For illustration, from inspection of figure 2 in \citet{Garraffo2018}, effects owing to magnetic field morphology may require the spin-down time to become reduced to $\sim0.1-1$\,Myr in some cases, during the first 100 million years. Accordingly, the effective spin-down timescale during early rotational evolution may drop to values capable of leading to significant misalignments even with correlation timescales of 1-10\,Myr or less.


\section{Summary}

In this work, we constructed a simple model to derive the dynamics of a star's spin axis as the mean angular momentum direction of its wind fluctuates. The physical scenario consisted of a star, losing angular momentum along an axis that is offset by a stochastically-varying small angle, of mean square value $\epsilon_0^2$, autocorrelated over a timescale $\tau_c$. Assuming that $\epsilon_0$ is small, most of the torque acts to decelerate the spin of the star over a timescale $\tau_{sp}$ that increases with time. 

Owing to the rapid decay of the spin-down timescale during the earliest phases of stellar evolution, most of the tilting typically occurs during the first 100\,Myr of stellar evolution (Figure~\ref{f5}). The magnitude of tilting depends entirely upon 2 parameters: $\epsilon_0$ and $\tau_{sp,0}/\tau_c$ (Equation~\ref{AnalyticSol}). 

Qualitatively, stellar obliquities of the same order as $\epsilon_0$ occur with high probability when the autocorrelation time exceeds the earliest spin-down time $\tau_{sp,0}$. From Figure~\ref{f5}, if $\tau_{sp,0}\sim 0.1\tau_{c}$ we see that in order to generate a Solar-like obliquity of 6\,degrees with 50\% probability, we require $\epsilon_0\gtrsim30\,^\circ$, however as $\tau_c$ gets large, the required $\epsilon_0$ reaches about 10\,degrees. 

The above quantitative results encapsulate two physical requirements that must be met if Solar-like stellar obliquities are to arise from stellar wind turbulence. First, tilting must occur over a spin-down time. If stochasticity averages out over shorter timescales (owing to a small $\tau_c$) than the spin-down timescale, then no matter how inclined the stellar wind's angular momentum axis is, there will typically be a small net deviation of orientation from the star's primordial position. 

The second, and more significant condition, is that tilting must occur early-on, i.e., within the first few 10s of millions of years after the dispersal of the protoplanetary nebula. It is possible to occur later, however, one must then postulate a scenario where the wind is coherently offset in a particular direction over timescales exceeding 10s-100s of millions of years. Accordingly, the key conclusion of this work is that wind-driven Solar obliquity is most likely excited soon after disk dispersal, making it a predicted feature of stars as young as 100\,Myr. We now discuss the feasibility of a misaligned, slowly-varying wind direction during the first 10-100 million years of stellar evolution.

\section{Discussion}

In the quantitative treatment above, we concluded that the solar wind must carry angular momentum away from the Sun at an angle $\epsilon_0\sim10^\circ$ and maintain that offset over timescales that compare to the spin-down timescale. The modern day Sun possesses a multi-Gyr spin-down timescale \citep{Metcalfe2019}, however, as argued above, younger stars exhibit much reduced spin-down timescales. A simple, Skumanich-like projection backwards in time would suggest spin down-times immediately after the dispersal of the protoplanetary disk on the order of 10\,Myr. Is it feasible that turbulence or stochasticity within the young Solar wind may exhibit coherence over such multi-million year timescales? To get some insight, the best place to look is the modern day Solar wind structure.

\subsection{Modern-day solar wind offset}

The Solar wind today causes an expansion of the Sun's magnetic field lines along the magnetic equator, generating a heliocentric current sheet (HCS; \citealt{Smith2001}), defined as the 2-dimensional region where the polarity of the magnetic field reverses. Owing to the offset between the Sun's spin axis and magnetic pole, this current sheet is not confined to one plane, but possess a waved structure as the field lines are forced into spirals by the Sun's rotation. Accordingly, in an instantaneous sense, the structure of the Solar wind is indeed tilted with respect to the Sun's rotation axis. Furthermore, the degree to which the magnetic dipole is offset, and thus the tilt angle of the HCS, changes during the Sun's 11-year Solar cycle.

Despite the HCS's tilt, the stellar dipole is locked into approximate corotation with the Sun. Accordingly, any tilting torques acting over a Solar rotation period are likely to become averaged out, owing to the relatively longer timescale of the Solar cycle. However, it is not clear whether tilting torques will average out over multiple Solar cycles. Specifically, in addition to the HCS's tilt, the Solar magnetic field exhibits a significant hemispheric asymmetry \citep{Oliver1994} that varies irregularly between successive Solar cycles. Thus at the end of one Solar cycle, the global topology of the Sun's magnetic field does not return exactly to the configuration that it had possessed at the beginning of that cycle.

Hemispheric asymmetries in the Solar dynamo are likely to arise in part from time-variability in the relative strengths of quadrupolar and dipolar field components \citep{Sokoloff1994}. In turn, magnetic topology directly influences the magnitude of magnetic braking \citep{Garraffo2016,Garraffo2018}. It remains unclear whether cycle-to-cycle variations in tilting torques truly cancel out, or whether there may exist a coherent sequence of torques acting over many cycles as a result of long-term topological variations. This latter case may lead to a greater degree of tilting then was computed in our idealized, dipolar approximation modeled above.

Variations in the solar wind will likely contribute to angular momentum loss only if they occur within the Alfv\'en radius. Unfortunately, in situ measurements from within the Alfv\'en radius are lacking, though this is likely to change owing to upcoming results from the Parker Solar Probe \citep{Fox2016}. Moreover even once a detailed picture of the Sun's close-in wind structure is obtained, it will remain difficult to ascertain how well this architecture may be applied to the young Sun. In the next subsection, we discuss some of aspects that may differ, however, for now it remains to be seen whether a long timescale offset exists between the Sun's spin axis and its net modern angular momentum loss.

Measuring the magnitude of the wind's offset may help constrain $\epsilon_0$ in the models above, however, it has only been possible to measure the Solar wind's variability over decadal timescales, or even millennial timescales via proxies such as radiocarbon measurements \citep{Beer2000,Bonev2004}. Such measurements may help in proposing reasonable values for the correlation time $\tau_c$. However, a million-year timescale record of Solar wind properties is not available, particularly not during the early life of the Solar system. 

It should be noted that the Earth's magnetic field does indeed undergo stochastic reversals, and that the power spectrum of this variability occurs with a 1/$f$ spectrum (where $f$ is frequency; \citealt{Pelletier2002}). The analogous reversals in stellar fields occur over decadal timescales \citep{Newkirk1980,Brown2011}, but interestingly, the Solar wind exhibits a similar 1/$f$ power spectral density over timescales ranging from days to years \citep{Kiyani2015}. The origin of such a spectrum is poorly understood, making its extrapolation to longer, million-year timescales uncertain. Nevertheless, if analogous to the geodynamo, longer-time correlations may exist in the Solar variability, providing power in timescales required for tilting torques to take effect.

Computational limitations prevent million-year timescale simulations of 3-D stellar winds \citep{Cohen2014,Cohen2017,Fionnagain2018}. Moreover, observational difficulties hinder the precise characterization of young stellar winds and astrospheres \citep{Wood2014}. Accordingly, we await future work in order to deduce the potential for an internal mode of variability to occur over the required timescales to significantly alter stellar orientations. 

\subsection{Early stellar environment}

Whereas it will be possible to learn a lot about the current Solar environment over the coming decade, the most important conlcusion of our modelling is that tilting must occur early-on. This is simply because the rate of tilting scales inversely with the spin-down timescale of the star ($\tau_{sp}$; equation~\ref{generalBeta}), which is largest during the earlier times \citep{Skumanich1972,Bouvier2013,Bouvier2014,Garraffo2018,Fionnagain2018}. 

As stated earlier, sophisticated 3-dimensional stellar models rarely examine the angular momentum components directed perpendicular to the stellar spin axis \citep{Holst2014,Garraffo2016,Cohen2017}. Furthermore, models used to examine likely wind properties of other stars, though calibrated using the Sun, struggle to reproduce Solar wind values in detail (see discussion in \citealt{Cohen2017}).

Given the above uncertainties, it is unlikely that MHD modeling would faithfully capture the long-term variability of off-axis angular momentum transport driven by internal modes of variability in young stellar winds. However, even if such modeling was accomplished, there is little reason to expect stochastic fluctuations autocorrelated over 10s of millions of years from internal modes alone. Nevertheless, the astrospheres of young stars are not shaped entirely by the stellar wind, but also by their interaction with the interstellar medium, as we now discuss.

\subsubsection{ISM ram pressure}
Tilting does not necessarily require that the variability in stellar winds is driven by internal fluctuations, only that the wind carries angular momentum off-axis. This misalignment may potentially be driven by interactions with dense gas in the young star's birth cluster environment \citep{Adams2010}. Interestingly, the typical relaxation times of open clusters span 10-100\,Myr \citep{Binney1987,Adams2010}. This timescale corresponds to the timescale over which our results suggest that a young star star may significantly alter its mean trajectory in the cluster (Figure~\ref{f5}). The star's trajectory would determine the direction of the oncoming ISM ram pressure, and in turn, the favored stellar wind direction\footnote{In the most extreme case, at least one supernova is widely considered to have supplied the early Solar system with short-lived radionuclides \citep{Adams2010}--suggesting that the blast waves thereof may have further altered the wind's structure.}. Accordingly, we hypothesize that the gaseous environment of a young, open cluster may reasonably lead to a stochastic forcing, that is coherent over sufficient timescales to induce stellar obliquities.

Today, the Solar wind's shape within a few AU of the Sun is well-described as an azimuthally-symmetric expansion of plasma, at velocity $v_{sw}$ and density $\rho_{sw}$ \citep{Parker1965}. However, on a larger scale (10s of AU), the on-coming ram pressure of the interstellar medium sculpts the heliosphere into a highly asymmetric, tear-drop shape, with the direction of relative motion roughly aligned with the local interstellar magnetic field direction \citep{Schwadron2014}.

 At a sufficiently large heliocentric distance, the Solar wind will equal the pressure exerted by the ISM \citep{Parker1965,Holzer1989,Zank1999}, at which point a termination shock forms. Both Voyager spacecraft have passed by this termination shock, with Voyager 1 indicating its distance is 94\,AU and Voyager 2 at 84\,AU \citep{Stone2008}. The fact that these two distances differ by about 10\% indicates that even today the shape of the heliosphere is significantly warped by the oncoming ISM. However, the modern-day Alfv\'en radius lies orders of magnitude closer to the Sun than the termination shock (within $\sim0.1$\,AU). Therefore, the asymmetry in the termination shock is most likely inconsequential for angular momentum loss from the Sun today, given that most of the angular momentum transfer between the Sun and the wind occurs within the Alfv\'en radius.

Whereas the ISM pressure has little influence upon the Sun's spin axis today, the size of the heliosphere is a dynamic quantity, changing signficiantly as the ISM pressure varies \citep{Muller2006}. Considering the simplest, spherically-symmetric model of the Solar wind, the ram pressure $\rho_{sw}v_{sw}^2$ scales roughly as distance $r^{-2}$ \citep{Parker1965}. At present, the ISM number density is roughly $n_{\textrm{ISM}}\equiv n_{\textrm{ISM},0}\sim0.2$\,cm$^{-3}$, but this value varies by many orders of magnitude over time \citep{Wyman2013,Schrijver2016}. 
 
 In its most general sense, let us suppose that the termination shock distance $r_{TS}$ is given by a balance of ISM pressure and wind ram pressure $(\rho_{sw}v_{sw}^2\propto n_{ISM})$, such that
 \begin{align}
 r_{TS}=r_{TS,0}\bigg(\frac{n_{ISM}}{n_{ISM,0}}\bigg)^{-1/2},
\end{align}
where $r_{TS,0}$ and $n_{ISM,0}$ are present-day values. High densities in the interstellar medium may readily shrink the termination shock to 10s of AU, or even less if we consider the conditions under which the young Solar system formed \citep{Adams2010}. Accordingly, this would impinge significantly upon the general architecture of the Solar wind. Related, observed features are the C-shaped jets emanated from young stars in the Orion nebula \citep{Canto1995,Bally2006}, which indicate the direct influence of the ISM upon the young star's outflows. However, this sculpting would only have a significant effect upon angular momentum loss if it was somehow communicated to the Sun via Alv\'en waves. The Alfv\'en point corresponds to where the wind becomes super-Alfv\'enic and thus the directionality imposed by the ISM must extend, at least approximately, to the Alfv\'en surface.
 
 Number densities exceeding 100 times that of the current value persist throughout the galaxy, and in star-forming regions analogous to the Sun's birth environment such as the Orion Nebula. Here, densities may rise even higher--up to 10$^{4}$ times modern \citep{Odell2001}. These higher densities have the potential to heavily sculpt the astrospheres over length-scales that become comparable to 1\,AU, but this remains significantly larger than the estimated modern Solar Alfv\'en radius, which is only about 0.1\,AU.
 
Even the modern Alfv\'en radius remains uncertain, and so combined with the young Sun's suspected increased in magnetic field, spin rate and mass-loss rate, a precise model of the Alfv\'en surface is difficult to reconstruct \citep{Garraffo2016,Garraffo2018}. Nevertheless, simple approximations exist, both from analytical calculations \citep{Lovelace2008} and comparisons to numerical models \citep{Matt2008}, which facilitate a general analysis of whether the Alfv\'en radius is expected to be significantly larger or smaller during these early times. Here, we utilize the approximation outlined in \citet{Lovelace2008}, where the Alfv\'en radius takes the form (in cgs units)
\begin{align}
r_A(t)&\approx\bigg(\frac{3}{2}\bigg)^{1/2}\bigg[\frac{B_\star^2 R_\star^4}{\Omega_\star \dot{M}_\star}\bigg]^{1/3}\nonumber\\
&\approx5\,\textrm{AU}\bigg(\frac{P_\star}{1\,\textrm{day}}\bigg)^{\frac{1}{3}}\bigg(\frac{B_\star}{1\,\textrm{kG}}\bigg)^{\frac{2}{3}}\bigg(\frac{R_\star}{R_\odot}\bigg)^{\frac{4}{3}}\bigg(\frac{\dot{M}_\star}{\dot{M}_\odot}\bigg)^{-\frac{1}{3}},
\end{align}
where $B_\star$ is the surface field strength of the star, $\dot{M}_\odot=2\times10^{-14}M_\odot$yr$^{-1}$, is the modern-day Solar wind magnitude and $P_\star$ is the spin period of the star\footnote{This analytic form predicts $r_A\approx0.15\,$AU today, which is slightly higher than some recent estimates, see e.g. \citet{DeForest2014}.}. In assuming larger mass-loss more sophisticated MHD models \citep{Vidotto2013} have predicted smaller $r_A$, consistent with the trend suggested by the expression above. Accordingly, it remains an important open question to determine the early mass-loss rates of stars \citep{Wood2014}.

For the sake of completeness we may compute an approximate value for $\tau_{sp,0}$ using relationship~\ref{tausp} using the fiducial parameters entered for $r_A$ above. Using $k\approx0.2$ and $\mu$ of order unity, $\tau_{sp,0}\sim10^6$\,years, which is consistent with expectations, though as previously mentioned, the scenario is too complex to compute spin-down timescales from first principles using such simplified expressions.

The appropriate parameters for the young Sun are highly uncertain and variable. For example, the magnitude of $\dot{M}_\star$ may vary potentially 4 orders of magnitude \citep{Drake2013,Fionnagain2018}, and the stellar radius of young stars is typically larger than the modern solar radius \citep{Shu1987}. Nevertheless, to an order of magnitude, the expression above suggests that the Alfv\'en radius may extend to distances comparable to an AU, while the termination shock may be pushed inward to a similar region. Accordingly, stochasticity inherent to the Sun's birth environment has the potential to directly impinge upon its angular momentum loss, contributing to asymmetric torques that could tilt the star. 

Our proposal that the ISM pressure influenced the young Sun's spin direction may be tested using more sophisticated modeling in the future, but shall be left has a hypothesis in this work. A problem that may arise from this framework lies in the fact that if indeed the termination shock becomes comparable to the Alfv\'en radius, the wind may never become super-Alfv\'enic, which is assumed in most existing stellar wind models. Future modeling shall require an incorporation of this alteration in order to first investigate whether the ISM is capable of causing long-term tilts in the stellar wind angular momentum, and second, whether this translates to off-axis torques upon the stellar spin axis.

\begin{figure}[h]
\centering
\includegraphics[trim=0cm 0cm 0cm 0cm, clip=true,width=1\columnwidth]{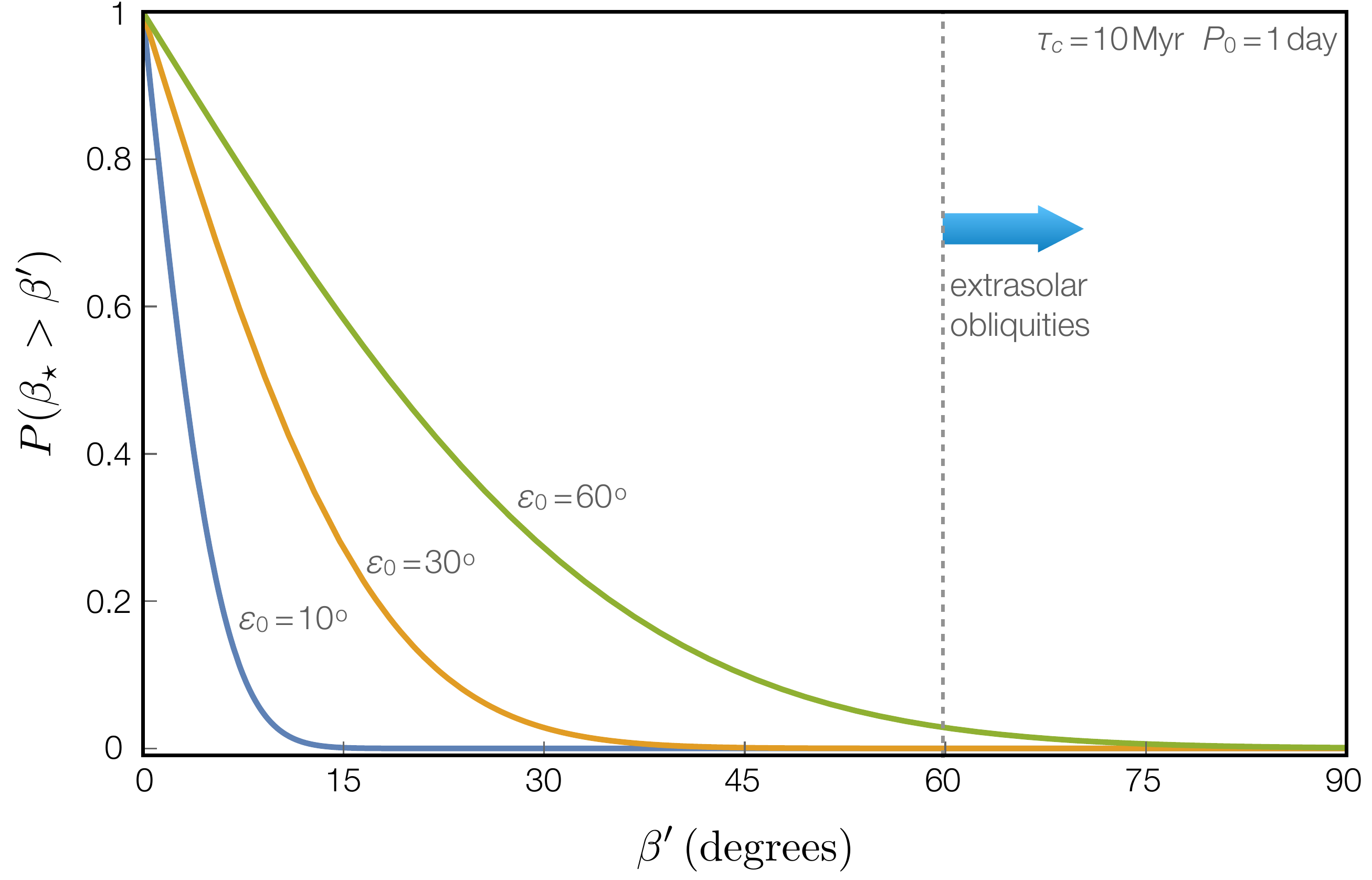}
\caption{The probability of stellar obliquity exceeding an angle $\beta'$ (Equation~\ref{PDF}) for 3 different values of stellar wind offset (blue line, $\epsilon_0=10^\circ$; orange line $\epsilon_0=30^\circ$; green line $\epsilon_0=60^\circ$). The small angle assumption begins to break down at the latter of these offsets, but nevertheless illustrates that extreme offsets are required to provide signifiant probabilities to generate extrasolar spin-orbit misalignments of 60\,degrees or more \citep{Winn2010}.}
\label{f7}
\end{figure}

\subsection{Relevance to exoplanetary systems}

The Solar obliquity is small relative to obliquities measured in many stars hosting extrasolar planets \citep{Winn2017}. These include examples of systems possessing the full range of misalignments, from aligned to anti-aligned \citep{Albrecht2012}. Could the stellar wind be responsible for these much more extreme misalignments? As a simple illustration, we plot in Figure~\ref{f7} the probability of stellar obliquity exceeding a value $\beta'$ (see Equation~\ref{PDF}) for $\epsilon_0=10^\circ$, $30^\circ$ and the more extreme $60^\circ$, choosing an optimistic case where $\tau_c=10t_0$. 

Even the most favourable stellar wind offsets are incapable of generating obliquities in excess of 60\,degrees more than $10\%$ of the time. Accordingly, whereas some rare instances might exist where stars become significantly tilted by stellar wind variability, it is unlikely to be a dominant driver of spin-orbit misalignments.

\section{Conclusions}

We proposed that the young Solar wind may have contributed to the $6^\circ$ obliquity possessed by the Sun. Stellar winds are known to decelerate the spins of Sun-like stars by an order of magnitude over time. Consequently, the Solar wind has removed a cumulative magnitude of angular momentum from the Sun that totals $\sim10$ times the current angular momentum possessed by the Sun. If, on average, the mean direction of this removed angular momentum differed by a few degrees, it would imply that the original spin axis of the Sun was off-set by a few degrees from its present location.

We constructed a stochastic model to determine the evolution of a stellar spin axis in response to a wind that stochastically varied its mean angular momentum axis by a typical angle $\epsilon_0$, autocorrelated with itself over a timescale $\tau_c$. We found that the modern Solar obliquity arises with high probability if $\epsilon_0\sim10^\circ$, and the autocorrelation time of turbulent fluctuations exceeds the spin-down time of the Sun during its first $\sim10$\,Myr. Quantitatively, if the Sun left the pre-main sequence phase with a spin period of 1\,day, and/or a spin-down timescale of 10\,Myr, turbulence autocorrelated with itself over 10\,Myr or more is required to induce significant obliquities. 

We argued that the required stellar wind variability is unlikely to arise purely from internal processes driving the Sun's wind--these occur over faster timescales and would be averaged out by the Sun's rapid rotation. Rather, we propose that stochastic interactions within the dense interstellar medium within the young birth cluster may warp the magnetosphere down to a few AU from the young Sun. Combined with the $10-100$\,Myr relaxation timescales of young open clusters, the oncoming ISM ram pressure would provide long-timescale fluctuations upon the stellar wind's orientation. Our hypothesized scenario would benefit from future, magnetohydrodynamic simulations of the interaction between a young star's magnetosphere, wind and the rapidly-evolving cluster within which it formed. 

A stars's birth cluster may play an under-appreciated role in shaping the resulting planetary systems that form around it. For example, a nearby supernova is thought to have injected short-lived radionuclides to our early Solar System \citep{Adams2010}, companion stars may torque entir protoplanetary disks out of alignment with extrasolar hosts \citep{Spalding2014a}, and the dispersal of planet-forming disks may be hastened by the radiation fields of nearby stars \citep{Owen2011}. If the Sun's own wind tilted its axis, steered by its birth environment, then the Solar obliquity stands as an additional, 4.5 billion year old relic of the formation environment of our Solar System.

\begin{acknowledgments}
 C.S thanks Gregory Laughlin, Ofer Cohen, Fred Adams and Konstantin Batygin for useful discussions and comments, along with the 51 Pegasi b Heising-Simons Foundation grant for their generous support. We also thank the referee, for in-depth comments that contributed to a substantially improved manuscript.
\end{acknowledgments}

\end{document}